Novel solid state vacuum quartz encapsulated growth of p-Terphenyl: the parent High $T_c$ Oraganic Superconductor (HTOS)

P.K. Maheshwari, R.S. Meena, Bhasker Gahtori, Reena Goyal, Rabia Sultana, Poonam Rani, and V.P.S. Awana[#]

*CSIR- National Physical Laboratory, Dr. K. S. Krishnan Marg, New Delhi-110012, India*

We report an easy and versatile route for the synthesis of the parent phase of newest superconducting wonder material i.e. p-Terphenyl. Doped p-terphenyl has recently shown superconductivity with transition temperature as high as 120K. For crystal growth, the commercially available p-Terphenyl powder is pelletized, encapsulated in evacuated ($10^{-4}$ Torr) quartz tube and subjected to high temperature (260C) melt followed by slow cooling at 5C/hour. Simple temperature controlled heating furnace is used during the process. The obtained crystal is one piece, shiny and plate like. Single crystal surface XRD (X-ray Diffraction) showed unidirectional (00*l*) lines, indicating that the crystal is grown along c-direction. Powder XRD of the specimen showed that as grown p-Terphenyl is crystallized in monoclinic structure with space group $P2_1/a$ space group, having lattice parameters a = 8.08(2)Å, b = 5.62(5)Å and c= 13.58(3)Å. Scanning electron microscopy (SEM) pictures of the crystal showed clear layered slab like growth without any visible contamination from oxygen. Characteristic reported Raman active modes related to C-C-C bending, C-H bending, C-C stretching and C-H stretching vibrations are seen clearly for the studied p-Terphenyl crystal. The physical properties of crystal are yet underway. The short letter reports an easy and versatile crystal growth method for obtaining quality p-terphenyl. The same growth method may probably be applied to doped p-terphenyl and to subsequently achieve superconductivity to the tune of as high 120K for the newest superconductivity wonder i.e., High $T_c$ Oraganic Superconductor (HTOS).

Key words: High $T_c$ Oraganic Superconductor (HTOS), p-Terphenyl, Crystal Growth, Structural details and Raman Spectroscopy

PACS number(s): 74.70.-b, 74.25.F-, 74.10.+v, 74.25.Ha

**[*]Corresponding Author**
Dr. V. P. S. Awana: E-mail: awana@mail.nplindia.org
Ph. +91-11-45609357, Fax-+91-11-45609310
Homepage awanavps.webs.com

Superconductivity often quoted as little above 100 years young; keep on adding new members and repeatedly proving the fact time and again that yes the phenomenon is as young as was a century before. The most recent addition is K doped p-Terphenyl with transition temperature as high as 123K [1,2]. Superconductivity is supposed to reside in one dimensional C-C chains of crystalline $K_3C_{18}H_{14}$ phase [3]. Though some other experimental reports [4,5] appeared soon, supplemented by theoretical efforts [6,7], yet the old hype like the one for infamous High $T_c$ Cuprates [8], $MgB_2$ [9], or Fe pinctides/chalcegonides [10,11] is somehow missing. Perhaps, the reason behind the missing old hype is related to the synthesis and growth of pure/K doped p-Terphenyl single crystals. Complicated chemical processes are reported for growth of K doped superconducting p-Terphenyl [1-5]. On the other hand in literature, the single p-Terphenyl is grown from its melt by Bridgman technique as well [12]. Both the chemical route [1-5] and the advanced Bridgman technique [12], hinders common researchers to achieve the newest wonder superconducting material i.e., K doped p-Terphenyl, nick named as High $T_c$ Oraganic Superconductor (HTOS).

Keeping in view, the fact that though an exciting new member i.e., K doped p-Terphenyl has joined the exotic superconductors community and yet the usual race is missing, the current short note opens new avenue to obtain the material by a simple solid state synthesis route. We report here a novel but simple synthesis route for the growth of quality p-Terphenyl. The commercially available p-Terphenyl powder is pelletized, encapsulated in evacuated ($10^{-4}$Torr) quartz tube and subjected to high temperature (260C) melt followed by slow cooling at 5C/hour. Interestingly, usual automated temperature controlled heating furnace is used during the process. We obtained shiny, one piece, transparent and plate like single crystal of the material. The XRD and SEM of the crystal proved the material to be of reasonably good quality. The short note will hopefully ignite the scientific community at large to race for the 123K superconductivity in K doped p-Terphenyl.

Commercially available Alpha Aesar p-Terphenyl with 99+ purity is ground properly with agate mortar and pelletized in rectangular form in argon filled M-Braun Glove Box. The pellet is subsequently put in a quartz tube, sealed gently at open end by cotton and rushed to glass blowing unit. The open end of the quartz tube is attached to a turbo pump, evacuated within minutes to $10^{-4}$ Torr and sealed immediately. The use of glove box and proper sealing of the evacuated quartz tube are necessary to avoid the oxygen contamination. The sealed quartz tube containing the p-Terphenyl pellet is now subjected to the heat treatment in furnace room. The sample is heated from room temperature to 170C at the rate of 8C/hour, and then to 260C at the rate of 3C/hour and hold at the same temperature for 12 hours. The sample is subsequently cooled from 260C to room temperature at the rate of 5C/hour. The schematic of the heat treatment is given in Fig. 1. X-ray diffraction (XRD) of the as obtained single crystalline material on its surface and in its powder form is done using Rigaku X-Ray Diffractometer. Scanning electron microscopy (SEM) pictures have been taken on ZEISS-EVO-10 electron microscope to understand the morphology of p-Terphenyl single crystal. Raman spectrum of p-Terphenyl single crystal is taken at room temperature using the Renishaw Raman Spectrometer.

Fig. 2 depicts the photograph of the as obtained p-Terphenyl single crystal. In fact the, whole pellet being sealed in evacuated quartz tube and subjected to heat treatment did grow in one piece (around 8cm) thin slab. The photograph being shown in Fig. 2, is only one piece from the same slab. We were bit worried before breaking the quartz tube, that may be crystal reacts immediately with open atmosphere. However we found that the crystal was stable even in open atmosphere. The crystal was soon subjected to Scanning Electron Microscopy (SEM) studies and the obtained picture is shown in Fig. 3. Clearly the morphology is slab like, showing unidirectional growth. The compositional analysis was also done and only negligible contamination of oxygen was seen. The exact compositional formula cannot be given, because of the limitations of the technique to account for the presence of Hydrogen.

A piece of the crystal was subjected to surface XRD and the result is shown in Fig. 4. It is evident from Fig. 4 that only (00*l*) planes are seen. This corroborates the SEM results (Fig.3) that the growth in slab like unidirectional in crystallographic c-plane. Subsequently, the crystal pieces are gently powdered and the PXRD (Powder XRD) was done at room temperature. The PXRD result is shown in Fig. 5 in 2Ѳ range of 10 to 50 Degree. The result shown in Fig. 5 is in good agreement to that as shown in ref. 12. This clearly shows that we have been able to grow p-Terphenyl single crystal by an easy and versatile method. The lattice parameters are lattice parameters a = 8.08(2)Å, b = 5.62(5)Å and c= 13.58(3)Å. We also measured the conductivity of the obtained p-Terphenyl single crystal and found the same to be not conducting at room temperature.

Figure 6 depicts the Raman spectra of p-Terphenyl single crystal done at room temperature using the Renishaw Raman Spectrometer. Here, the spectra were examined using a laser source having excitation photon energy of 2.41eV (514nm). Clear Raman peaks related t0 C-C-C bending, C-H bending, C-C stretching and C-H stretching vibrations are seen, which are comparable to the earlier reported result [12]. It is very clear from Fig. 5 that studied p-Terphenyl single crystal is of reasonably good quality. Various other physical property characterizations are in process and the trials for obtaining K doped p-Terphenyl superconducting single crystals are also underway.

Summarily, the current short note reports the crystal growth of p-Terphenyl via an easy and versatile synthesis route. The same synthesis route can be applied for K doped p-Terphenyl, which is reported [1] 123K superconductor. The sole aim of the short note is to ignite the race among solid state experimentalists to hunt for the newest wonder superconductor i.e., K doped p-Terphenyl. The same is already nick named as High $T_c$ Oraganic Superconductor (HTOS).

This work is financially supported by DAE-SRC outstanding investigator award scheme on search for new superconductors. Mrs. Shaveta Sharma is acknowledged for Raman spectrum on our p-Terphenyl crystal. Authors thank Prof. G. Baskaran from Inst. Mat. Sci. Chennai for persuading them to try to get the wonder new superconductor i.e., doped p-Terphenyl. Further studies for doping p-Terphenyl are underway and we are hopeful to establish the novel High $T_c$ Oraganic Superconductor (HTOS).

**Figure Captions:**

**Figure 1:** Flow chart of p-Terphenyl single crystal growth process.

**Figure 2:** Photograph of as obtained p-Terphenyl single crystal.

**Figure 3:** SEM images of p-Terphenyl single crystal.

**Figure 4:** Room temperature XRD pattern of p-Terphenyl single crystal.

**Figure 5:** Powder XRD pattern of crushed p-Terphenyl single crystal

**Figure 5:** Raman Spectrum of p-Terphenyl single crystal

Fig. 1

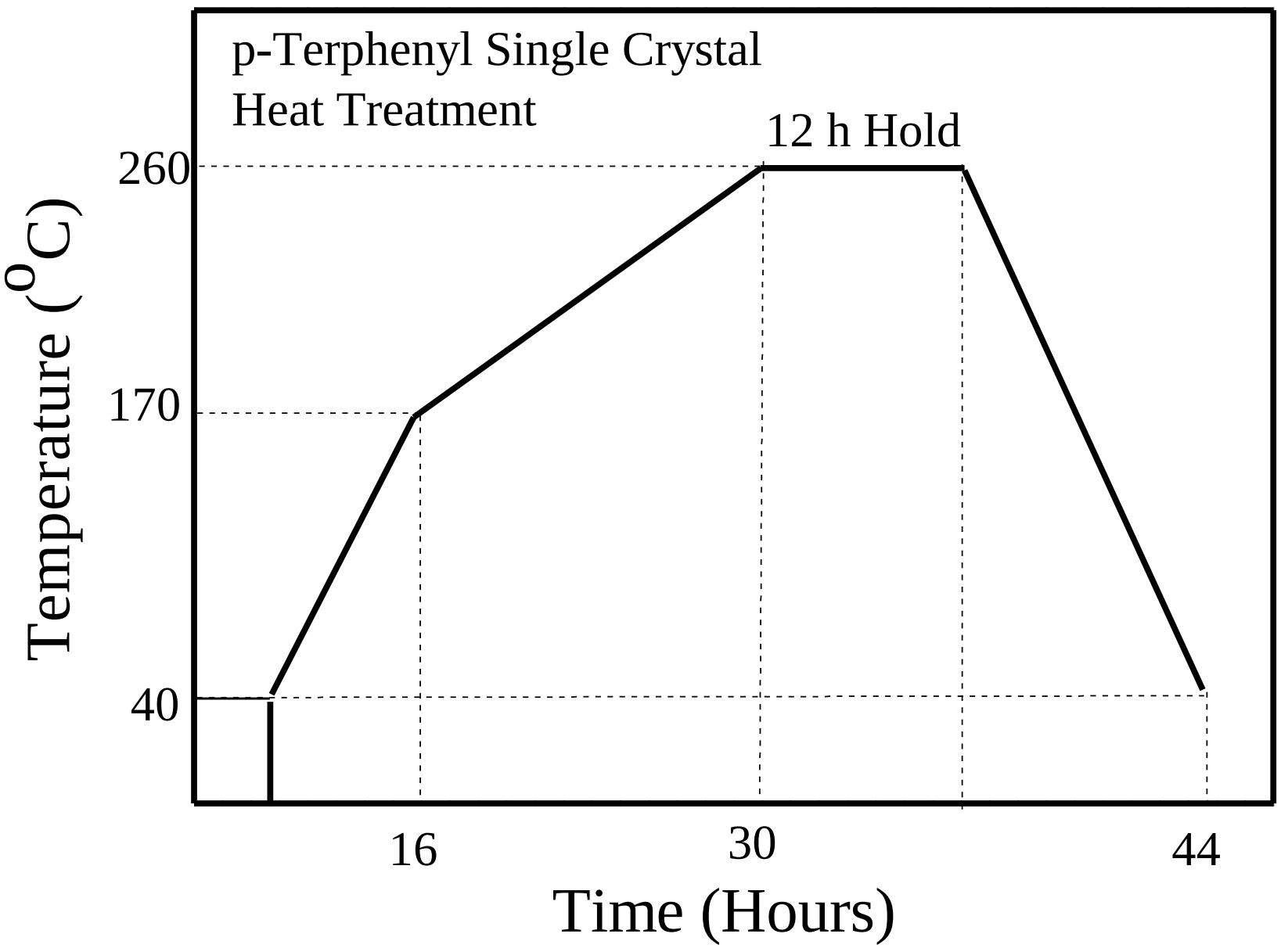

Fig. 2

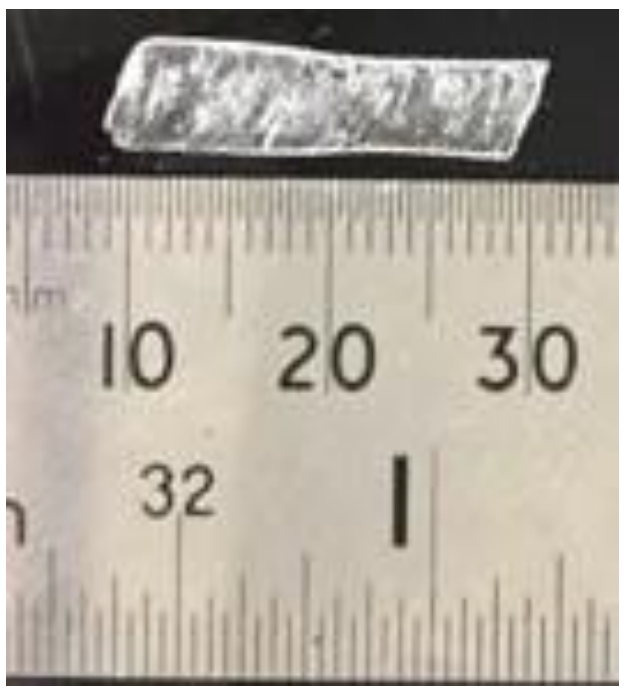

Fig. 3

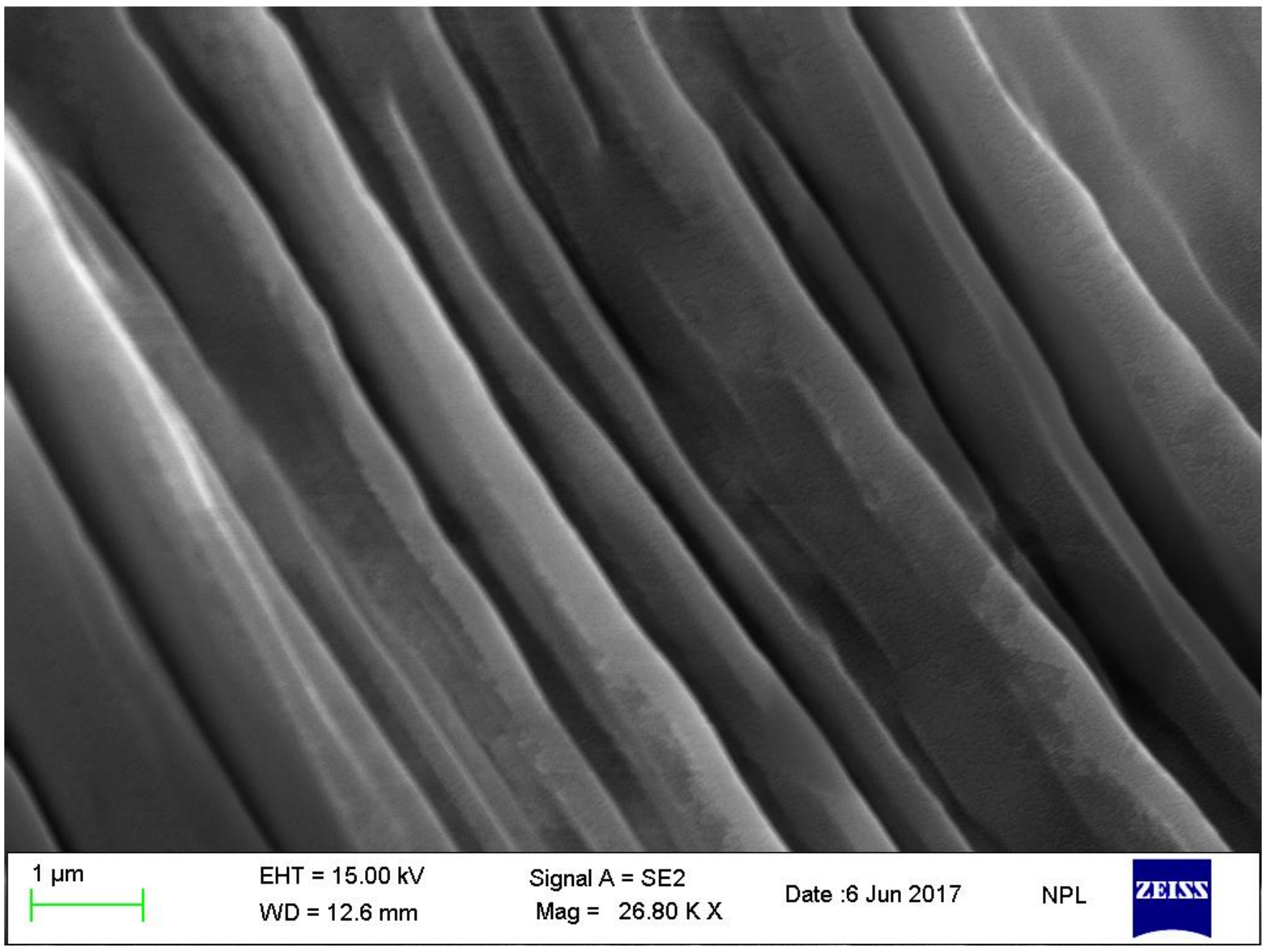
1 μm
EHT = 15.00 kV
WD = 12.6 mm
Signal A = SE2
Mag = 26.80 K X
Date :6 Jun 2017
NPL
ZEISS

Fig. 4

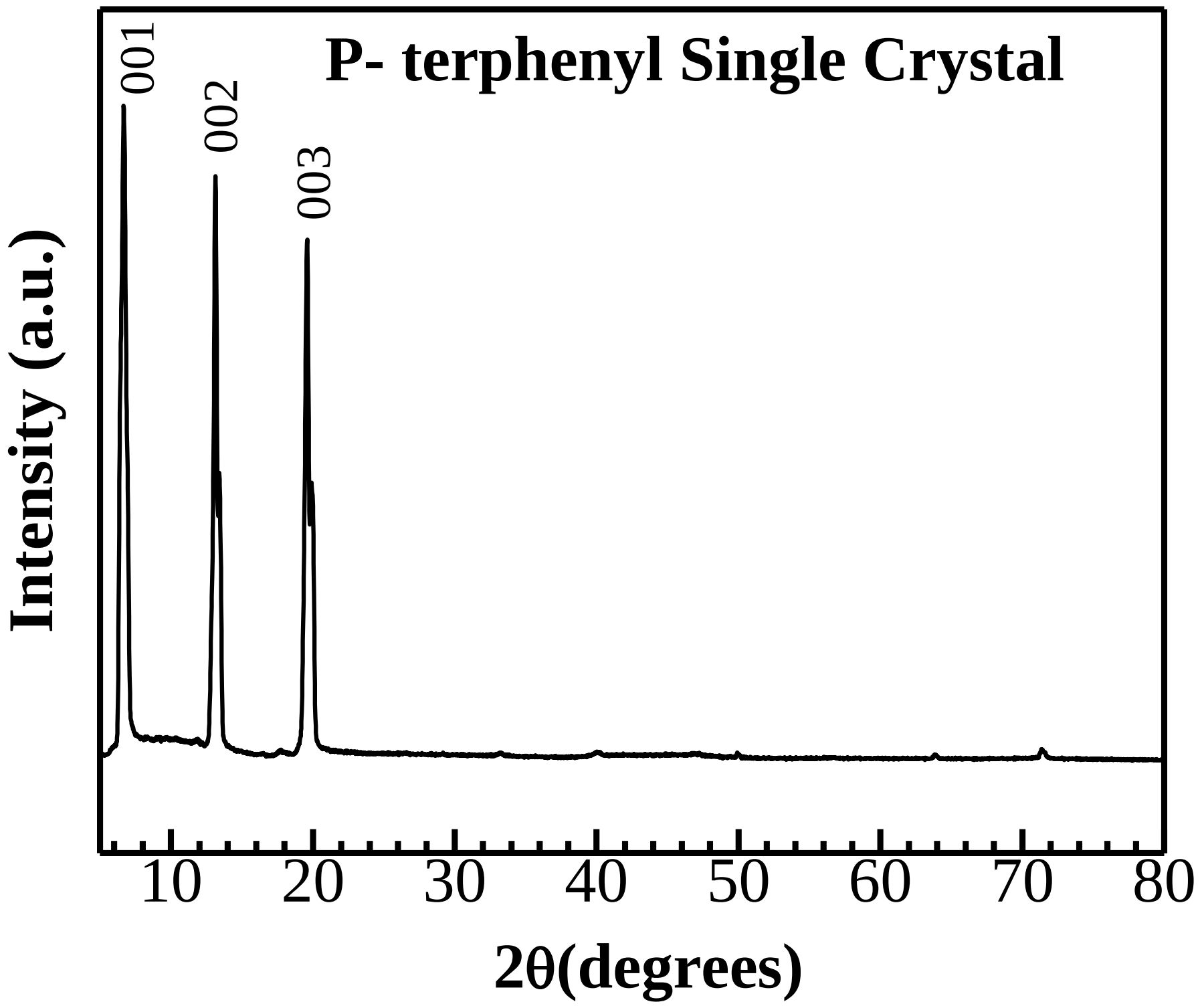

Fig. 5

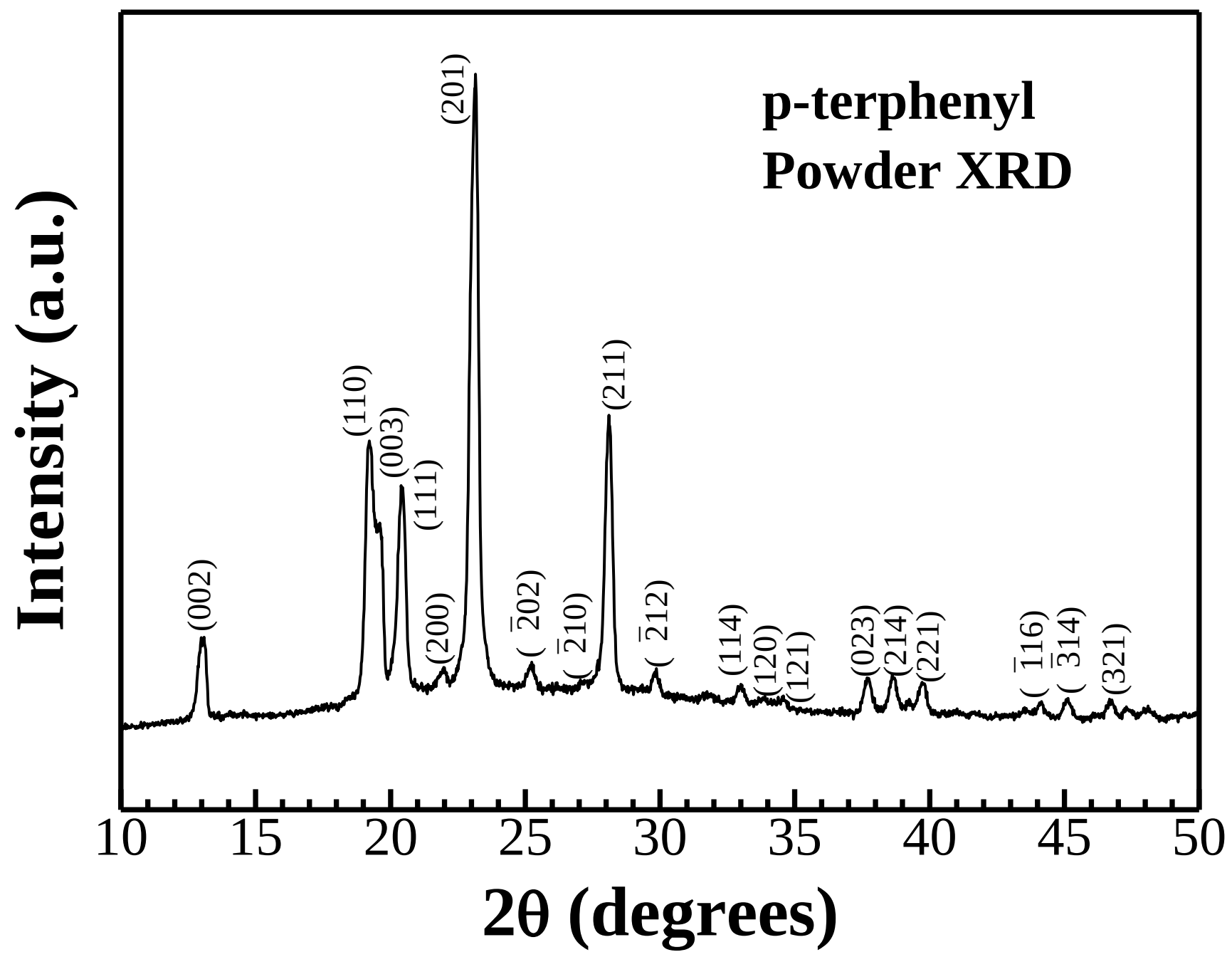

Fig. 6

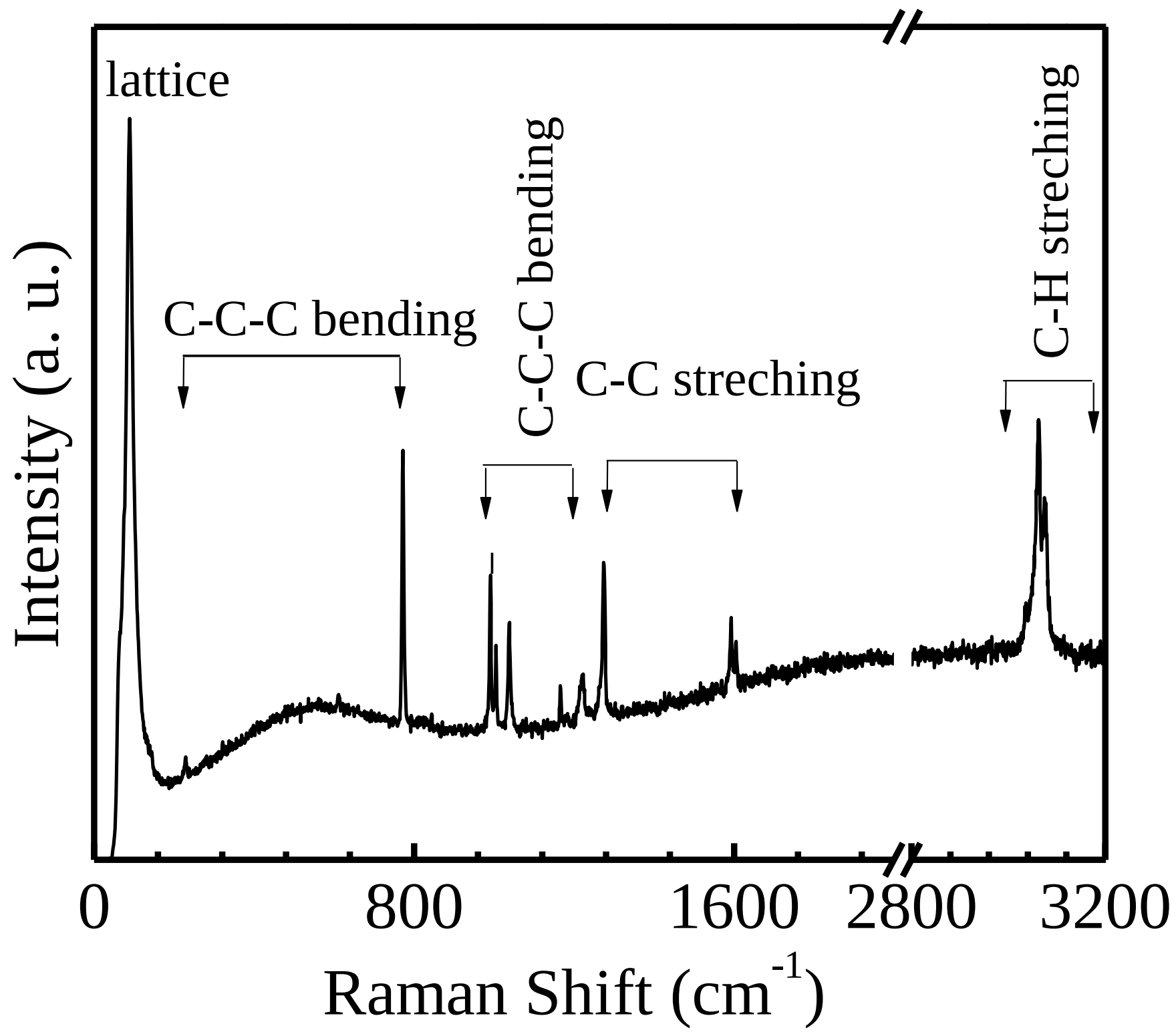